\documentclass[twocolumn, superscriptaddress,showpacs,preprintnumbers,amsmath,amssymb]{revtex4}
\usepackage{graphicx,tabularx}
\usepackage{bm}
\usepackage{epsfig}
\newcommand{\beq}{\begin{equation}}
\newcommand{\eeq}{\end{equation}}
\newcommand{\beqn}{\begin{eqnarray}}
\newcommand{\eeqn}{\end{eqnarray}}

\begin{document}


\title{Power-law in a gauge-invariant cut-off regularisation}

\author{T.~Varin, J.~Welzel, A.~Deandrea, D.~Davesne}

\affiliation{Universit\'e de Lyon, Villeurbanne, F-69622, France; Universit\'e Lyon 1,
Institut de Physique Nucl\'eaire de Lyon, 4 rue Enrico Fermi, 
F-69622 Villeurbanne, France
}


\begin{abstract}
We study one-loop quantum corrections of a compactified Abelian $5d$ 
gauge field theory. We use a cut-off regularisation procedure which respects the symmetries of the model, {\it i.e.} gauge
invariance, exhibits the expected power-like divergences and therefore allows 
the derivation of power-law behavior of the effective 4d gauge coupling in a
coherent manner. 
\end{abstract}

\maketitle

\section{Introduction}
Since a decade, it has been realized that large and/or universal 
extra-dimensions could be included in models beyond the Standard Model (SM) 
without being in conflict with experimental data~\cite{XdTeV,LED,RadiusLimit}. 
The SM can then be thought as a low-energy limit of a higher-dimensional theory. 
This kind of scenario offers new insights to challenging problems, like gauge 
hierarchy problem~\cite{LED,RS}, supersymmetry breaking~\cite{SUSYB}, 
electroweak symmetry breaking~\cite{EWSB}, dark matter~\cite{UED}, {\it etc}. 

Higher-dimensional gauge field theories are non-renormalisable and by
dimensional analysis, power-like divergences will appear in the loop 
integrations. A well-known feature of these theories is the power-like quantum 
corrections of effective four-dimensional gauge couplings~\cite{DDG}. 
For the purpose of computing loop corrections, it is of fundamental 
importance to respect the symmetries of the theory. On the one hand, dimensional 
regularisation~\cite{'tHooft:1972fi} is well-known to preserve gauge invariance 
but hides power-like divergences. Extensions to extra-dimensional theories are 
given in~\cite{extdim}. On the other hand, a naive proper-time cut-off 
regularisation provides the expected power-like divergences but breaks gauge 
symmetry (see for example~\cite{Zinn-Justin:2002ru}). Other approaches, 
preserving the symmetries of the theory are possible, for 
example Pauli-Villars regularisation (see \cite{Contino:2001gz} for an 
application to extra-dimensions). 

In this paper we study a compactified five-dimensional 
Abelian gauge theory with a cut-off regularisation which preserves 
gauge invariance and exhibits power-like divergences (details and applications 
to other subjects are given in~\cite{Cutoff}). We present briefly the model
in section II and introduce the regularisation to study Ward-identities in 
section III. We then focus on the calculation of the vacuum polarization 
function, section IV, to deduce the power-law behavior of the 
effective four-dimensional gauge coupling with respect to the cut-off, section
V. We end with a discussion on the results.

\section{A brief presentation of the model}
The action of quantum electrodynamics (QED) in $5$ dimensions, or a generic 5d 
Abelian gauge theory, is :
\beq
\mathcal{S}=\int 
d^5x\,\left(-\frac{1}{4}F^{MN}F_{MN}+\overline{\Psi}(i\gamma^MD_M-m_e)\Psi\right
)\label{5Daction}
\eeq
with the capital indices $M,N=(0,1,2,3,5)=(\mu,5)$. The gamma 
matrices are $\gamma^M=(\gamma^{\mu},\,i\gamma^5)$ and the five-dimensional 
covariant derivative is defined by : 
$$D_M=(\partial_{\mu}-i\tilde{e}A_{\mu},\,\partial_{5}-i\tilde{e}A_{5}).$$
The compactification on a circle $S^1$ implies the following Fourier decomposition 
of the fields:
\beqn
\Psi(x^{M})&=&\frac{1}{\sqrt{2\pi R}}\sum_{n=-\infty}^{+\infty}\psi^{(n)}
(x^{\mu})e^{inx^5/R}\\
A_{\mu}(x^{M})&=&\frac{1}{\sqrt{2\pi R}}
\sum_{n=-\infty}^{+\infty}A_{\mu}^{(n)}(x^{\mu})e^{inx^5/R}\\
A_5(x^{M})&=&
\frac{1}{\sqrt{2\pi 
R}}\sum_{n=-\infty}^{+\infty}A_{5}^{(n)}(x^{\mu})e^{inx^5/R}\; .
\eeqn

The four-dimensional fields $\psi^{(n)}$, $A_{\mu}^{(n)}$, $A_{5}^{(n)}$ are the 
Kaluza-Klein (KK) excitations (or modes) of the original five-dimensional 
fields $\Psi$, $A_{\mu}$ and $A_{5}$ respectively. 
Other compactifications are possible and indeed widely studied in the 
literature, depending on the precise field content and masses one wants to 
obtain in the four-dimensional theory. In the following we shall consider 
$S^1$, but the method can be easily applied to other cases (if one introduces
an additional $Z_2$ symmetry for example). 

Performing the integration over the extra-coordinate $x^5$ in eq.~(\ref{5Daction}) 
leads to the effective 4d action :
\beqn
\mathcal{S}_{4}
&=&\sum_{n=-\infty}^{+\infty}\int 
d^4x\,-\frac{1}{4}F^{\mu\nu (-n)}F^{(n)}_{\mu\nu}\nonumber\\
&+&\sum_{n=-\infty}^{+\infty}\int d^4x\,
\overline{\psi}^{(-n)}(i\not{\partial}-M_n)\psi^{(n)}\nonumber\\
&+&\sum_{n,m=-
\infty}^{+\infty}\int 
d^4x\,e\,\overline{\psi}^{(-n)}\gamma^{\mu}A_{\mu}^{(n-m)}\psi^{(m)}\nonumber\\
&+&\sum_{n,m=-\infty}^{+\infty}\int d^4x\,ie\,
\overline{\psi}^{(-n)}\gamma^{5}A_{5}^{(n-m)}\psi^{(m)}\nonumber\\
&+&\sum_{n=-\infty}^{+\infty}\int 
d^4x\,-\frac{1}{2}\partial_{\mu}A_5^{(-n)}\partial^{\mu}A_5^{(n)}\nonumber\\
&+&\sum_{n=-\infty}^{+\infty}\int d^4x\, 
\frac{-in}{R}A^{\mu (-n)}\partial_{\mu}A_5^{(n)}\nonumber\\
&+&\sum_{n=-\infty}^{+\infty}\int d^4x\, 
-\frac{1}{2}\frac{n^2}{R^2}A_{\mu}^{(-n)}A^{\mu (n)}\nonumber\\&+&\int 
d^4x\,~\mathcal{L}_{gauge-fixing}
\eeqn
The effective four-dimensional gauge coupling and the mass matrices of the
fermions are respectively 

$\displaystyle e=\frac{\tilde{e}}{\sqrt{2\pi R}}$ and $M_n=m_e + i\gamma^5n/R$. 
\par\hfill\par
The $4d$ sub-Lagrangian with zero-mode bosons and $n$ fermions is
invariant under the following $4d$ $U(1)$ gauge transformations :
\beqn
A_{\mu}^{(0)}&\to& A_{\mu}^{(0)}+\frac{1}{e}\partial_{\mu}\theta(x^{\mu})\nonumber \\
A_{5}^{(0)}&\to& A_{5}^{(0)}\nonumber\\
\psi^{(n)}&\to&e^{i\theta(x^{\mu})}\psi^{(n)}
\eeqn
This sector is the relevant one for the purpose of this paper. It will be
shown that the $U(1)$ gauge symmetry is preserved in the regularisation procedure.
The gauge-fixing term of this sector is taken to be the usual St\"uckelberg Lagrangian: 
\beq
\mathcal{L}_{gauge-fixing}=-\frac{\lambda}{2}(\partial^{\mu}A_{\mu}^{(0)})^2\; .
\label{GF}
\eeq
We will limit our study to renormalisation in the effective four-dimensional theory 
obtained after compactification. For a comparison of the one-loop renormalisation 
of the full extra  dimensional theory with the four-dimensional effective 
one see \cite{alvarez}.

\section{Regularisation and Ward identities}
Since we want to generate explicitly the high-energy dependence of 
the correlations functions at one-loop in our extra-dimensional model, we have 
to choose a cut-off regularisation procedure that fulfills the necessity of 
preserving the $U(1)$ gauge symmetry. Such procedure has been developed in detail 
in~\cite{Cutoff}. The strategy is to deduce all the integrals encountered during 
the regularisation procedure from only one single integral. More precisely, 
starting from~:
\begin{equation}
I(\alpha,\beta) \equiv \int 
\frac{d^dk}{i(2\pi)^d} \frac{1}{[\alpha k^2-\beta m^2]} 
\label{I_alpha_beta}
\end{equation}
we can deduce all the integrals of the general 
form (obtained after partial traces on gamma matrices and introduction of 
Feynman parameters)~:
\begin{equation}
\int \frac{d^dk}{i(2\pi)^d}\frac{k^a}{[k^2-m^2]^b}
\end{equation}
by derivations of (\ref{I_alpha_beta}) with respect to $\alpha$ and $\beta$.

The starting integral can be computed with the Schwinger proper-time method (for 
example, see \cite{Zinn-Justin:2002ru}), and reads~:
\begin{equation}
I(1,1)_{div}= - \frac{1}{4 \pi^2} (\Lambda^2 - m^2 \log \Lambda^2)
\end{equation}
where $\Lambda$ is the cut-off. The identification of the cutoff with a physical 
mass scale and the running with respect to the scale are discussed in detail 
in \cite{effscalepower}.

It is worthwhile to mention that a key-point for the consistency of the method 
is to take a special care of the dimension $d$, which has be taken equal to $4$ (resp. 
$2$) for logarithmic (resp. quadratic) terms (for further details, 
see~\cite{Cutoff}). 

The vacuum polarization function of the $4d$ 
photon, $A_{\mu}^{(0)}$, is the infinite sum of 
vacuum polarization in which massive Kaluza-Klein excitations run in the 
loop :
\begin{equation}
i\Pi^{\mu \nu} (p) = - e^2 \sum_{n=-\infty}^{+\infty}\int  
\frac{d^4k}{(2\pi)^4}Tr \left[ \gamma^\mu \frac{1}{k \hbox{\hskip -0.2 true 
cm}\slash -m_n} \gamma^\nu \frac{1}{k \hbox{\hskip -0.2 true cm}\slash - p 
\hbox{\hskip -0.2 true cm}\slash -m_n}\right]
\label{QED_vac}
\end{equation}
\begin{center}
\includegraphics*[scale=.9]{QED.epsi}\label{QEDa}
\end{center}
\vspace{3mm}
Neglecting for the moment the sum over the Kaluza-Klein modes, 
eq.(\ref{QED_vac}) is then formally equivalent to the standard 4-dimensional QED 
for a fermion of mass $m_n$. Thus, the polarization tensor 
reads~:
\begin{equation}
i\Pi^{\mu \nu} (p) = - 4 e^2 \int  \frac{d^4k}{(2\pi)^4} 
\frac{N^{\mu\nu}}{(k^2-m_n^2)((k-p)^2-m_n^2)}\label{QED_vac2}
\end{equation}
with
\begin{equation*}
N^{\mu\nu}=k^{\mu} (k^{\nu}-p^{\nu}) + k^{\nu} (k^{\mu}-p^{\mu}) -g^{\mu\nu} k(k-p) + g^{\mu\nu} m_n^2
\end{equation*}
After the regularisation procedure, the divergent part of $\Pi^{\mu \nu} (p)$ is~:
\begin{equation}
\Pi^{\mu \nu}_{div} (p) =- \frac{e^2}{12 \pi^2} (p^2 g^{\mu \nu} - p^{\mu} 
p^{\nu} ) \ln \biggl( \frac{\Lambda^2}{m_n^2} \biggr)
\end{equation}
As it should be, $\Pi^{\mu \nu}_{div} (p)$ is transverse (independently of the 
value of the Kaluza-Klein number $n$ in the loop). Moreover, it behaves 
logarithmically, in total agreement with what is expected for gauge theories. 

A complementary test of our regularisation procedure is the fact 
that the Ward identity between 
the three-point function $\Gamma _\mu (p,p)$  and the $n^{th}$ `electron' self-energy 
$\Sigma (p)$ is satisfied. This test can also be worked out in standard $4d$ QED
using our regularisation procedure. In our effective 
model, one has to check that the two above quantities, depicted diagrammatically 
below (we consider here the diagrams for the zero-mode photon $A_\mu^{(0)}$ 
stemming from the effective four-dimensional action we have described in section 
2), satisfy the following relation~:
\begin{equation}\Gamma _\mu (p,p) = - \frac{\partial}{\partial p 
^\mu}\Sigma(p)\label{Ward_id}
\end{equation}
\\
where the contributions to $\Gamma _\mu (p,p)$ can be decomposed into two parts 
$\Gamma^{(1)}_{\mu} (p,p)$ and $\Gamma^{(2)}_\mu (p,p)$, respectively 
represented by:\\
\begin{tabular}{m{4.2cm} m{4.2cm}}
\hspace{5mm}
\includegraphics*[scale=.75]{Ward.epsi}
\label{QED}
&
\hspace{5mm}
\includegraphics*[scale=.75]{Ward3.epsi}
\label{QED2}\\
\end{tabular}\\
Likewise, the contributions $\Sigma^{(1)} (p)$ and $\Sigma^{(2)} (p)$ to the 
$n^{th}$ mode `electron' self-energy are:\\
\begin{tabular}{m{4.5cm} m{4cm}}
\hspace{5mm}
\includegraphics*[scale=.9]{Ward2.epsi}
\label{QED3}
&
\includegraphics*[scale=.9]{Ward4.epsi}
\label{QED4}\\
\end{tabular}

Following our regularisation procedure, it is straightforward to obtain~:
\begin{eqnarray}
-i \Sigma^{(1)} (p) &=& \frac{i(-i e)^2}{16 \pi^2} \log \Lambda ^2 \left( 
(\frac{1}{\lambda}+3)M_n - \frac{1}{\lambda} p \hbox{\hskip -0.2 true cm}\slash 
\right)\nonumber\\-i \Sigma^{(2)} (p) &=& \frac{i (i e^2)}{32 \pi^2} \log 
\Lambda^2 \left( 2 M_n - p \hbox{\hskip -0.2 true cm}\slash\right)\nonumber\\-i e 
\Gamma^{(1)} _\mu(p,p)&=&\frac{-(-i e)^3}{16 \pi^2 \lambda}~\gamma _\mu \log 
\Lambda ^2 \nonumber\\-i e \Gamma^{(2)} _\mu(p,p)&=&\frac{-(-i e)^3}{32 
\pi^2}~\gamma _\mu \log \Lambda ^2  
\end{eqnarray}
where $\lambda$ is the parameter of the St\"uckelberg gauge fixing term,
eq. (\ref{GF}). We see explicitly that (\ref{Ward_id}) is satisfied, the 
gauge-invariance of the sector with the zero-mode
$A_{\mu}^{(0)}$ is preserved.

\section{Vacuum polarization}

The aim of this section is to apply our regularisation procedure to the
vacuum polarization function of $A_{\mu}^{(0)}$ in the effective theory obtained 
by compactifying on $S^1$ the original $5$-dimensional QED Lagrangian. Taking 
into account the sum over the Kaluza-Klein modes, one obtains~:
\begin{equation}
\Pi^{\mu \nu}_{div} (p) = -\sum_{n=-\infty}^{\infty} \frac{e^2}{12 \pi^2} (p^2 
g^{\mu \nu} - p^{\mu} p^{\nu} ) \ln \biggl( \frac{\Lambda^2}{m_e^2+n^2/R^2} 
\biggr)
\end{equation}
Our strategy is then to separate the standard $4$-dimensional part ($n = 0$) 
from the extra dimension contributions ($n \neq 0$). Then, we have:
\begin{equation}
\begin{split}
\ln\biggl(\prod_{n=-\infty}^{\infty}\frac{\Lambda^2}{m_e^2+n^2/R^2}\biggr) =
\ln\biggl(\frac{\Lambda^2}{m_e^2}\biggr) \hspace{2cm}\\ \hspace{5mm}+ 2 
\ln\biggl(\prod_{n=1}^{\infty}\frac{\Lambda^2}{m_e^2+n^2/R^2}\biggr)
\end{split}
\label{log}
\end{equation}
Some approximations can be done in order to simplify the calculation of the previous expression
(for an alternative analytical approach see the appendix of \cite{Ghilencea:2003xy})~:
\begin{enumerate}
\item 
$m_e^2 \ll 1/R^2$\\
This can be justified by the fact that we assume the first Kaluza-Klein 
resonance to be far above the fundamental mass, which is the case in
phenomenological applications~\cite{RadiusLimit}. 
\item 
$n_{max} = \Lambda R \gg 1$\\
In the spirit of a Wilsonian effective theory~\cite{Georgi}, $\Lambda$ is the typical scale 
that enables to select the relevant degrees of freedom present in the theory~: 
thus the sum over $n$ is truncated at some value $n_{max} \simeq \Lambda R$. 
Moreover since we are interested in the high-energy behavior of the theory, we 
will make the assumption $n_{max} \gg 1$ in the following. 

The limits of the truncation of KK sums have been discussed by 
Ghilencea~\cite{Ghilencea}. It has been shown that, when one performs 
the infinite KK sums, higher derivative operators of higher dimension are 
generated as one-loop counterterms describing a non-decoupling effect of the very 
massive KK states. However, this effect appears only in six dimensions 
or when one sums over 2 KK numbers, which is not our case.\end{enumerate}
Using these two approximations, we can then rewrite the second term of eq. 
(\ref{log}) as~:
\begin{equation}
\ln\biggl(\prod_{n=1}^{n_{max}} \frac{\Lambda^2 R^2}{n^2}\biggr) = \ln\biggl(\frac{(\Lambda^2
R^2)^{\Lambda R}}{(\Lambda R)!^2}\biggr) \sim 2 \Lambda R-\ln\Lambda R
\end{equation}
where the Stirling formula has been used in the last step.

Finally, the total contribution for the divergent part of the vacuum polarization function reads~:
\begin{equation}
\Pi^{\mu \nu}_{div} (p) = -\frac{e^2}{12 \pi^2} (p^2 
g^{\mu \nu} - p^{\mu} p^{\nu} ) (4 \Lambda R)
\end{equation}
In agreement with the results obtained in the literature~\cite{DDG}, the 
divergence is linear in the cut-off. We see by the preceding manipulations 
that the power-law is the result of the sum of the individual logarithmic 
contributions.

\section{Renormalisation of the $4d$ effective coupling constant}
\begin{figure}[htbp]
\begin{center}
  \mbox{\epsfxsize=0.45\textwidth
       \epsffile{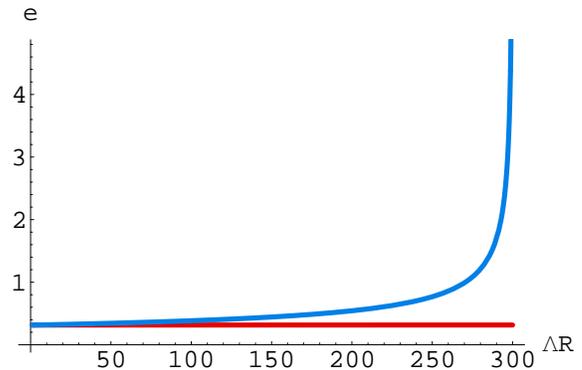}}
  \end{center}
\caption[]{\label{PowerLawFig} Cut-off dependence of $e$ in the model discussed
(upper curve) and in the standard $4d$ QED (lower curve), as a function of
$\Lambda R$. The radius is arbitrarily chosen at (100 GeV)$^{-1}$. The Landau pole
is at $\Lambda =300\,R^{-1}$. }
\end{figure}

In an Abelian theory, due to Ward identities, the beta function of the 4d 
effective coupling $e$ can be calculated directly :
\beq
\beta_e=-\frac{e}{2}\Lambda\frac{\partial\Pi_{\Lambda}}{\partial\Lambda}
\eeq
where $\Pi_{\Lambda}$ is the divergent part of scalar vacuum polarization 
function defined by :
\beq
\Pi_{\Lambda}=\frac{1}{3}g_{\mu \nu}\Pi^{\mu
\nu}_{div}=\frac{-e^2}{3\pi^2}\Lambda R
\eeq
We obtain :
\beq
\beta_e=\frac{e^3}{6\pi^2}\Lambda R
\eeq
which gives the following asymptotic behavior of $e(\Lambda)$ with respect to 
the cut-off, between the scale $\mu=R^{-1}$ and $\mu=\Lambda$ :
\beq
e(\Lambda)=\left(\frac{e(R^{-1})^2}{1-\frac{e(R^{-1})^2}{3\pi^2}(\Lambda
R-1)}\right)^{1/2}
\eeq
This expression shows the expected power-law running of $e$,
figure~(\ref{PowerLawFig}). It admits a Landau pole at
$\displaystyle\Lambda_L=(1+\frac{3\pi^2}{e(R^{-1})^2})R^{-1}.$

Equivalently, we can write :
\beq
\alpha_e^{-1}(\Lambda)=\alpha_e^{-1}(R^{-1})-\frac{b}{2\pi}X(\Lambda R-1)
\eeq
with $\alpha_e=\frac{e^2}{4\pi}$, $b=4/3$ and $X=2$. We found the same result
for $\alpha^{-1}(\Lambda)$ as in~\cite{DDG} but 
with the fundamental difference that we do not need any final rescaling of the 
cut-off.

\section{Discussion and conclusions}
Regularisation dependence of quantum corrections in 
higher-dimensional field theory has been extensively discussed. It is of 
fundamental importance for all phenomenological applications such as the study 
of divergences in a given model (gauge couplings, yukawas etc). Indeed, at the 
end of the regularisation process, one would like to identify the cut-off 
$\Lambda$ with the mass scale $M$ at which our effective theory breaks down. 
However, the identification cannot be done with the knowledge of the effective 
theory only. One needs a matching with the theory taking place at $M$ to fix the 
coefficient of the power-like quantum corrections. For an example of this 
type of calculations see \cite{Contino:2001si}. 

Without any UV completion of the 
higher-dimensional model, one can fix the coefficient by an {\it external} 
requirement. In a sense, the choice of a given regularisation procedure (and the 
definition of the cut-off) is a feature of the model. For example, the authors 
of~\cite{Kobayashi}, asked the gauge couplings in the $4d$ effective theory to 
recover the result of the non-compactified theory in the limit of large radius. 
In the calculation of~\cite{DDG}, the cut-off was redefined in order to recover 
the asymptotic result of including KK states at their thresholds, between 
$\mu=\frac{n}{R}$ and $\mu=\frac{n+1}{R}$ and so on. As in all standard cut-off 
calculations, gauge invariance was explicitly broken through the appearance of 
a $\Lambda g_{\mu\nu}$ term with no compensating $p_{\mu}p_{\nu}$ in the vacuum
polarization $\Pi_{\mu\nu}$. 

In our calculation, we included all states of KK number $\mid n \mid\leq \Lambda 
R$ at once, assuming the decoupling of the heavy states and without 
introducing any {\it ad-hoc} redefinition of the cut-off. Moreover, we 
explicitly checked in section III that our calculation of loop integrals does not 
break gauge-invariance. The main advantage of the method proposed in this paper 
is the possibility to keep the physical insight of the cut-off procedure 
together with symmetry conservation.

\end{document}